\documentclass[aps,pra,10pt,twocolumn,showpacs,superscriptaddress,footnoteinbib]{revtex4}
\usepackage{amsmath}
\usepackage{latexsym}
\usepackage{amssymb}
\usepackage{color}
\usepackage{graphics,epstopdf}
\usepackage{soul}

\begin{document}

\title{ Preservation of quantum key rate in the presence of decoherence}

\author{Shounak Datta}
\email{shounak.datta@bose.res.in}
\affiliation{S. N. Bose National Centre for Basic Sciences, Block JD, Sector III, Salt Lake, Kolkata 700098, India}

\author{Suchetana Goswami}
\email{suchetana.goswami@bose.res.in}
\affiliation{S. N. Bose National Centre for Basic Sciences, Block JD, Sector III, Salt Lake, Kolkata 700098, India}

\author{Tanumoy Pramanik}
\email{tanu.pram99@bose.res.in}
\affiliation{S. N. Bose National Centre for Basic Sciences, Block JD, Sector III, Salt Lake, Kolkata 700098, India}

\author{A. S. Majumdar}
\email{archan@bose.res.in}
\affiliation{S. N. Bose National Centre for Basic Sciences, Block JD, Sector III, Salt Lake, Kolkata 700098, India}

\date{\today}

\begin{abstract}

It is well known that  the interaction of quantum systems
with the environment reduces the inherent quantum correlations. 
Under special circumstances the effect of decoherence can be reversed, for example, 
the interaction modeled by an amplitude damping channel can boost the 
teleportation fidelity from the classical to the quantum region for a bipartite 
quantum state. Here, we first show that this phenomena fails in the case of a quantum key
distribution protocol.  We further show that the technique of weak measurement
can be used to slow down the process of decoherence, thereby helping to preserve the 
quantum key rate when  one or both systems are interacting with the environment 
via an amplitude damping channel. Most interestingly, in certain cases weak measurement with
post-selection where one considers both success and failure of the technique
is shown to be more useful than without it when both systems interact with the environment.
\end{abstract}


\maketitle

\section{Introduction}

Correlations between quantum systems can not always be explained by local causal theory~\cite{EPR, Bell}. This nature of quantum correlations helps to perform certain information processing tasks, for example, quantum teleportation~\cite{Tele}, super dense coding~\cite{SDC} and quantum key distribution~\cite{QKD1, QKD2}, which are not possible using classical correlations.  
However,  in practice, quantum systems are continuously interacting with the environment, and 
this interaction weakens the correlations between observed quantum systems. Hence, the most 
crucial task  in quantum information processing is to protect quantum correlations from
diminishing due to the effect of the ubiquituous environment.

Under special circumstances, interaction between systems and a common environment can generate 
entanglement~\cite{ent_gen}. For example, when two or more atoms are consecutively passing 
through a cavity, they become entangled~\cite{ent_gen2, ent_gen3}. Although, for a specific 
information processing task, {\it viz.} quantum teleportation, the environmental interaction 
modeled by an amplitude damping channel (ADC) can enhance the fidelity of quantum teleportation 
of those bipartite states whose teleportation fidelity lies just below the quantum 
region~\cite{Fid_enhan}, this improvement of fidelity is found to be possible only for a 
certain class of bipartite states~\cite{Fid_Som, Fid_Tan}.

Moreover, one can use the technique of weak measurements to protect the fidelity of quantum 
teleportation when systems are interacting with the environment modeled by an amplitude damping 
channel~\cite{Fid_Tan, Weak_M_1, Weak_M_2, Weak_M_3, Weak_M_4, Weak_M_5, Weak_M_6}. The idea of
weak measurements was originally proposed~\cite{Weak_M_0} on the basis of weak coupling
between the observed system and the measurement device, thereby making possible for the 
measurement outcomes to be amplified compared to the eigenvalue spectrum of original system,
for suitable post-selected ensembles. This technique  has been implemented in many different 
ways, such as in the study of the
spin Hall effect~\cite{SHE}, superluminal propagation of light~\cite{SPL}, wave particle duality 
using cavity-QED experiments~\cite{WPD}, direct measurement of the quantum wave 
function~\cite{MQWF}, measurement of ultrasmall time delays of light~\cite{MUTDL}, and  
observing Bohmian trajectories of photons~\cite{TOP,pg}.

In the present work, we study the possibility of preservation of the quantum key rate for a
bipartite state shared between Alice and Bob where Alice's system is not trusted as a quantum 
system. More specifically, we discuss a way to protect the one-sided device independent quantum 
key distribution (1s-DIQKD) protocol~\cite{1sdiqkd} when the system interacts with the environment modeled by 
ADC. Comparing the preservation of 1s-DIQKD with the preservation of the fidelity of quantum 
teleportation, we observe that ADC cannot improve  the optimal secret key rate in 1s-DIQKD,
which is derived using the steering 
inequality~\cite{St_Fur} based on the fine-grained uncertainty relation~\cite{FUR_2}, though it 
can improve the teleportation fidelity for states having teleportation fidelity just below the 
quantum region~\cite{Fid_enhan, Fid_Som}. We show that improvement of the key rate becomes 
possible using the technique of weak measurement and its reversal, which may be used to 
suppress the effect of the amplitude damping 
decoherence~\cite{Weak_M_1, Weak_M_2, Weak_M_3, Weak_M_4, Weak_M_5, Weak_M_6}.

This paper is organized as follows. In Sec. II, we briefly recapitulate  the technique of weak 
measurement and its reversal in the presence of an interaction of the system with the 
environment as modeled by ADC. In Sec. III  we discuss the connection of steerability with 
quantum key distribution for the case of the 1s-DIQKD protocol~\citep{St_Fur}. In Sec. IV we 
demonstrate the effect of the amplitude damping decoherence on the steerability and key rate. 
In Sec. V we show how the technique of weak measurement and its reversal can be used to protect 
the key rate. Finally, in Sec. VI we summarize the main results of this work.

\section{Weak and reverse weak measurement in the presence of an amplitude damping channel}

Let us consider that a qubit is prepared either in the state $|0\rangle_S$  or in the state $|1\rangle_S$. The qubit is allowed to interact with the environment  by ADC, where the environment 
is initially in the state $|0\rangle_E$. Due to the effect of  decoherence, the combined state of the system and environment becomes
\begin{eqnarray}
|0\rangle_S~|0\rangle_E \, &\rightarrow & \, |0\rangle_S~|0\rangle_E \nonumber \\
|1\rangle_S~|0\rangle_E \, &\rightarrow & \, \sqrt{\overline{D}_S}\,|1\rangle_S~|0\rangle_E \, + \,\sqrt{D_S}\, |0\rangle_S~|1\rangle_E,
\label{ADC_int}
\end{eqnarray}
where $D_S$ is the strength of the system with environment and $\sqrt{\overline{D}_S}=1-D_S$. In practice, the photon loss when the photon is passing through the environment can be regarded as amplitude damping decoherence. The above interaction~(\ref{ADC_int}) can be written as a positive  and trace preserving map $\Lambda$ given by
\begin{eqnarray}
\Lambda(\rho) = W_{S, 0}\,\rho\,W_{S,0}^{\dagger} \, +\, W_{S, 1}\,\rho\,W_{S,1}^{\dagger},
\label{ADC_Map}
\end{eqnarray}
where 
\begin{eqnarray}
W_{S, 0} = \begin{pmatrix}
1 & 0\\
0 & \sqrt{\overline{D}_S}
\end{pmatrix}; ~~~~~~~~
W_{S,1}=\begin{pmatrix}
0 & \sqrt{D_S}\\
0 & 0
\end{pmatrix},
\label{KOP_ADC}
\end{eqnarray}
and $\sum_{i=0}^1\,W_{S,i}=I$.


It has been  shown in earlier works~\cite{Weak_M_1, Weak_M_2, Weak_M_3, Weak_M_4, Weak_M_5} 
that the technique of weak measurement and its reverse can suppress the environmental effect 
modeled by ADC. Here, before allowing interaction with the environment, the system is measured 
using a scheme of weak quantum measurement, with strength $p_S$. More specifically, the detector detects the system with probability $p_S$ if and only if the system is  in the state $|1\rangle_S$. When the detector detects, the correponding Kraus operator is given by
\begin{eqnarray}
M_{S,\,1}\,=\,\begin{pmatrix}
0 & 0\\
0 & \sqrt{p_S},
\end{pmatrix}
\label{Weak_Detec}
\end{eqnarray}
which does not have any inverse. Hence, this operation, i.e., the detection is irreversible. The operator when the system is not detected is given by
\begin{eqnarray}
M_{S,0}\,=\,\begin{pmatrix}
1 & 0\\
0 & \sqrt{\overline{p}_S}.
\end{pmatrix}
\label{Weak_Rev}
\end{eqnarray}
The operator $M_{S,0}$ is reversible, i.e., the application of its inverse restores the system to its initial state. The case where the system is detected will be discarded. Hence, weak measurement is associated with a success probability.

After performing weak measurement, the system is allowed to interact with the environment and at the end, to reduce the affect of environment, reverse weak measurement is performed. The 
operator corresponding the case when the system is not detected is given by
\begin{eqnarray}
N_{S,0}\,=\,\begin{pmatrix}
\sqrt{\overline{q}_S} & 0\\
0 & 1
\end{pmatrix},
\label{Rev_Weak}
\end{eqnarray}
where $q_S$ is the strength of the reverse weak measurement.

\section{Steering and its connection with 1s-DIQKD}

Non-local quantum correlations between two systems, say $A$ and $B$, can be categorized 
separately by entanglement, steering and Bell non-local correlation~\cite{Jones_07}, respectively. In the case of entanglement, both $A$ and $B$ are trusted as quantum systems, whereas, none 
of them is trusted as a quantum system in Bell non-local correlation. In the intermediate 
case of steering, one of them is trusted as a quantum system and the shared state, $\rho_{AB}$ 
is said to be entangled if it cannot be described by a local hidden state model (LHS)~\cite{Jones_07}. There are different steering criteria based on different uncertainty relations~\cite{St_Fur, St_Reid, St_Entropy, St_Exp}. In the present work we use the optimal fine-grained steering criteria to study the quantum key rate of steerable states~\cite{St_Fur}. 

To discuss fine-grained steering, let us consider the following game. Alice prepares a large 
number of bipartite quantum states $\rho_{AB}$. She then sends all the systems $B$ to Bob and 
keeps the systems $A$ with her. Bob only trusts that the system $B$ is quantum, but agrees that 
the prepared state is entangled if and only if Alice has control on the state of systems $B$. 
In other words, $\rho_{AB}$ is said to be steerable when it cannot be explained a by local 
hidden state model~\cite{Jones_07}. To check whether the state is steerable, Bob asks Alice to 
control the state of his system $B$ in one of the eigenstates of the observable chosen randomly 
from the set $\{\sigma_z,\,\sigma_x\}$. Next, Alice measures a suitable observable chosen from 
the set $\{\mathcal{A}_1,\,\mathcal{A}_2\}$ and communicates her choice and outcome. The shared 
state $\rho_{AB}$ is steerable when the conditional probability distribution $P(b_{\sigma_{z\,(x)}}|a_{\mathcal{A}_i})$ (where $b$ and $a$ are measurement outcomes at Bob's and Alice's side) violates 
the relation~\cite{St_Fur}
\begin{eqnarray}
\frac{1}{2}\left[P(b_{\sigma_z}|a_{\mathcal{A}_1}) + P(b_{\sigma_x}|a_{\mathcal{A}_2})\right] \leq \frac{3}{4} .
\label{St_FUR}
\end{eqnarray}

In Ref.~\cite{St_Fur}, it has been further shown that if the shared state $\rho_{AB}$ between 
systems $A$ and $B$ is maximally steerable, then none of these systems can be quantumly 
correlated, or steerable with any other system -- this phenomena is called monogamy of 
steerable states. This nature of steerable states lower bounds the secret key 
rate in a one-sided device independent way, i.e., one of the systems is not trusted. The 
lower bound of the secret key rate $r$, corresponding to  $\rho_{AB}$ which violates the 
above inequality~(\ref{St_FUR}) is given by~\cite{St_Fur}
\begin{eqnarray}
r \geq \log_2\left[ \frac{\frac{3}{4}+\delta}{\frac{3}{4}-\delta} \right],
\label{Key_Id}
\end{eqnarray}
where $\delta$ is the degree of violation of the inequality.

\section{Lower bound of quantum key rate under amplitude damping channel}

In the above section, we have discussed the connection of the secret key rate with steerability. 
 In the considered steering game, Alice needs to send the quantum system $B$ to Bob through 
the environment. In the derivation of secret key rate represented by inequality~(\ref{Key_Id}), the interaction between the system and the environment is not considered. In this section, we 
will study the effect of the environment on steerability and hence, on the key rate. Here, we 
discuss two different cases separately. In the first case, {\it ``Case-I"}, we consider the 
effect of environment on the system $B$ when it is passing through the environment. In the second
 case, {\it ``Case-II"}, we discuss the effect on the key rate when both systems interact with 
the environment through amplitude damping decoherence. In both the cases we assume that Alice \
prepares the systems $A$ and $B$ in one of the maximally entangled states given by
\begin{eqnarray}
|\psi^{\pm}\rangle\,&=&\, \frac{|00\rangle\,\pm\,|11\rangle}{\sqrt{2}},\nonumber \\
|\phi^{\pm}\rangle\,&=&\, \frac{|01\rangle\,\pm\,|10\rangle}{\sqrt{2}}.
\label{State_Ini}
\end{eqnarray}
 
{\it Case I.} Here we discuss the environmental effect on the steerability when the system $B$ interacts with environment via ADC during the time of its passage. After environmental interaction, the shared state between Alice's system $A$ and Bob's system $B$ becomes 
\begin{eqnarray}
\rho_{AB}^{\prime} &=& (I\otimes W_{2,0})\,|\psi^{\pm}\rangle\langle \psi^{\pm}|\,(I\otimes W^{\dagger}_{2,0})\nonumber \\
&& +\,(I\otimes W_{2,1})\,|\psi^{\pm}\rangle\langle \psi^{\pm}|\,(I\otimes W^{\dagger}_{2,1}), \nonumber\\
&=& \begin{pmatrix}
\frac{1}{2} & 0 & 0 & \pm \frac{\sqrt{1-D_2}}{2} \\
0 & 0 & 0 & 0 \\
0 & 0 & \frac{D_2}{2} & 0\\
\pm \frac{\sqrt{1-D_2}}{2} & 0 & 0 & \frac{1-D_2}{2}\\
\end{pmatrix},
\label{P_Env_B_WO}
\end{eqnarray}
when Alice  prepares the initial state $|\psi^{\pm}\rangle$ given by Eq.(\ref{State_Ini}), or 
\begin{eqnarray}
\sigma_{AB}^{\prime}&=& (I\otimes W_{2,0})\,|\phi^{\pm}\rangle\langle \phi^{\pm}|\,(I\otimes W^{\dagger}_{2,0})\nonumber \\
&& +\,(I\otimes W_{2,1})\,|\phi^{\pm}\rangle\langle \phi^{\pm}|\, (I\otimes W^{\dagger}_{2,1}), \nonumber \\
&=& \begin{pmatrix}
\frac{D_2}{2} & 0 & 0 & 0 \\
0 & \frac{1-D_2}{2} & \pm \frac{\sqrt{1-D_2}}{2} & 0\\
0 & \pm \frac{\sqrt{1-D_2}}{2} & \frac{1}{2} & 0\\
0 & 0 & 0 & 0\\
\end{pmatrix},
\label{AP_Env_B_Wo}
\end{eqnarray}
where Alice prepares the initial state $|\phi\rangle$ given by Eq.(\ref{State_Ini}). The Kraus operators $W_{2,0}$ and $W_{2,1}$ are given by Eq.(\ref{KOP_ADC}). The strength of the environmental interaction with the system $B$, $D_2$ lies in the range $0\,\leq\,D_2\,\leq\,1$. 

Next, we we discuss the steerability and the lower bound of the key rate of the state $\rho_{AB}^{\prime}$ ($\sigma_{AB}^{\prime}$).
We calculate the maximum value of the quantity $[(P(b_{\sigma_z}|a_{\mathcal{A}_1})\,+\,P(b_{\sigma_x})|a_{\mathcal{A}_2})/2]$ (left-hand side of Eq.(\ref{St_FUR})), where maximization is taken over Alice's 
choice of observables $\mathcal{A}_1$ corresponding to spin measurement along the direction $\hat{n}_1$,  and $\mathcal{A}_2$ along the direction $\hat{n}_2$.  For both the prepared 
states $|\psi\rangle^{\pm}$ and $|\phi\rangle^{\pm}$, the above quantity becomes
\begin{eqnarray}
\frac{1}{2}(P(b_{\sigma_z}|a_{\sigma_z})\,+\,P(b_{\sigma_x}|a_{\sigma_x})\, = \, \frac{3+\sqrt{1-D_2}}{4},
\label{St_CI}
\end{eqnarray}
where Alice's optimal measurement setting is spin measurement along the $z$-direction 
($x$-direction) when Bob measures along that direction, i.e., $\mathcal{A}_1=\sigma_z$ ($\mathcal{A}_2=\sigma_x$). The lower bound of the key rate when the system $B$ interacts with environment is thus given by
\begin{eqnarray}
r_B = \log_2\left[ \frac{3+\sqrt{1-D_2}}{3-\sqrt{1-D_2}}\right]
\label{K_CI}
\end{eqnarray}

\begin{figure}[!ht]
\resizebox{8cm}{5.5cm}{\includegraphics{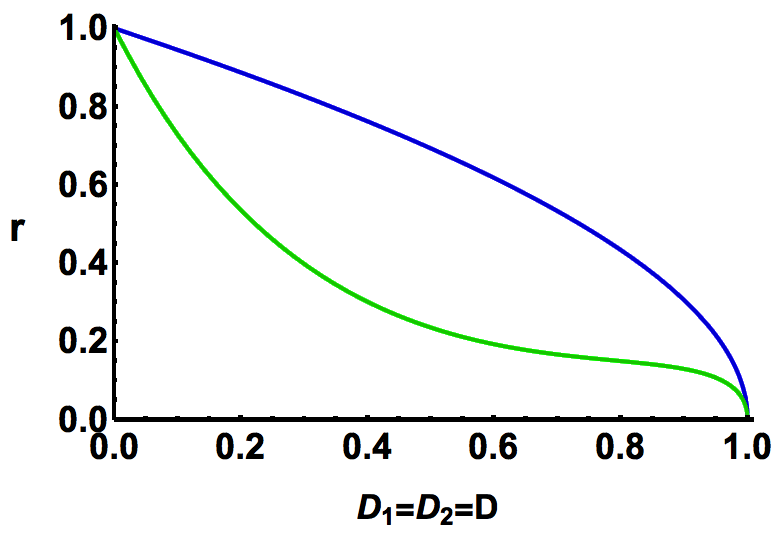}}
\caption{\footnotesize The lower bound of key rate for the initial state $|\psi^{\pm}\rangle$ given by Eq.(\ref{State_Ini}) is plotted versus the decoherence parameter $D$.  The upper curve is 
for the case when the system $B$ only is affected by decoherence, and lower curve is for the
case when both systems interact with the environment. 
}
\label{r_P_WOW}
\end{figure}

\begin{figure}[!ht]
\resizebox{8cm}{5.5cm}{\includegraphics{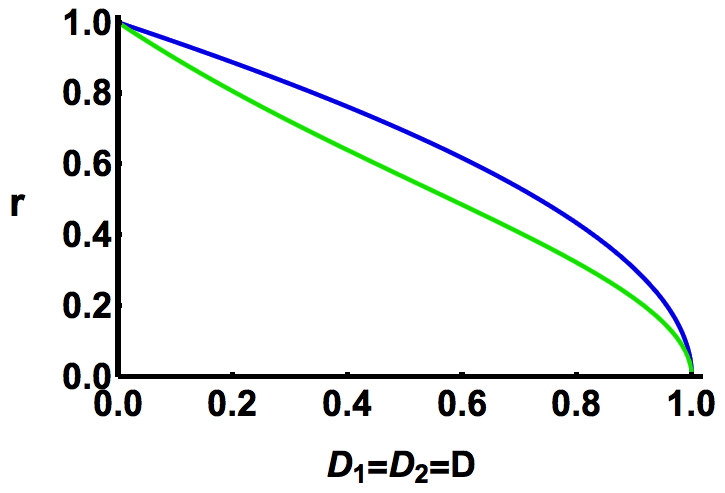}}
\caption{\footnotesize The lower bound of key rate for the initial state $|\phi^{\pm}\rangle$ given by Eq.(\ref{State_Ini}) is plotted versus the decoherence parameter $D$.  The upper curve is 
for the case when the system $B$ only is affected by decoherence, and lower curve is for the
case when both systems interact with the environment. 
}
\label{r_AP_WOW}
\end{figure}

{\it Case II.} Now we consider that both systems, $A$ and $B$ interact with environment under 
amplitude damping decoherence. After environmental interaction of both systems, the shared 
state becomes  either
\begin{eqnarray}
\rho_{AB}^{\prime\prime} &=& (W_{1,0}\otimes I)\,\rho_{AB}^{\prime}\,( W^{\dagger}_{1,0} \otimes I)\nonumber \\
&&+\,(W_{1,1} \otimes I)\,\rho_{AB}^{\prime}\,(W^{\dagger}_{1,1} \otimes I), \nonumber\\
&=&\begin{pmatrix}
\frac{1+D_1D_2}{2} & 0 & 0 & \pm \frac{\sqrt{\overline{D}_1 \overline{D}_2}}{2}\\
0 & \frac{\overline{D_1} D_2}{2} & 0 & 0\\
0 & 0 & \frac{D_1 \overline{D}_2}{2} & 0 \\
\pm \frac{\sqrt{\overline{D}_1 \overline{D}_2}}{2} & 0 & 0 \frac{\overline{D}_1 \overline{D}_2}{2}
\end{pmatrix}
\label{P_Env_AB_WO}
\end{eqnarray}
where $\rho_{AB}^{\prime}$ is given by Eq.(\ref{P_Env_B_WO}), or 
\begin{eqnarray}
\sigma_{AB}^{\prime\prime} &=& (W_{1,0}\otimes I)\,\sigma_{AB}^{\prime}\,( W^{\dagger}_{1,0} \otimes I)\nonumber \\
&&+\,(W_{1,1} \otimes I)\,\sigma_{AB}^{\prime}\,(W^{\dagger}_{1,1} \otimes I), \nonumber\\
&=& \begin{pmatrix}
\frac{D_1+D_2}{2} & 0 & 0 & 0\\
0 & \frac{\overline{D}_1}{2} & \pm \frac{\sqrt{\overline{D}_1 \overline{D}_2}}{2} & 0\\
0 & \pm \frac{\sqrt{\overline{D}_1 \overline{D}_2}}{2} & \frac{\overline{D}_2}{2} & 0\\
0 & 0 & 0 & 0\\
\end{pmatrix}
\label{AP_Env_AB_WO}
\end{eqnarray}
where $\sigma_{AB}^{\prime}$ is given by Eq.(\ref{AP_Env_B_Wo}). Again, for both the 
states $\rho_{AB}^{\prime \prime}$ and $\sigma_{AB}^{\prime \prime}$, the optimal set of measurement 
settings for Alice is $\{\mathcal{A}_1=\sigma_z,\,\mathcal{A}_2=\sqrt{1-D^2}\,\sigma_x-D \sigma_z\}$, where we consider both systems $A$ and $B$ interact with the same environment, i.e., $D_1=D_2=D$. The left-hand side of Eq.(\ref{St_FUR}) becomes
\begin{eqnarray}
\frac{1}{2}\left[P(b_{\sigma_z}|a_{\mathcal{A}_1}) + P(b_{\sigma_x}|a_{\mathcal{A}_2})\right] = \frac{3+D + 2\, D^2+ \sqrt{1-D^2}}{4+4\,D},
\label{St_P_WOW}
\end{eqnarray}
when the shared state is $\rho_{AB}^{\prime \prime}$, and 
\begin{eqnarray}
\frac{1}{2}\left[P(b_{\sigma_z}|a_{\mathcal{A}_1}) + P(b_{\sigma_x}|a_{\mathcal{A}_2})\right] = \frac{3+3\,D + \sqrt{1-D^2}}{4+4\,D},
\label{St_AP_WOW}
\end{eqnarray}
when the shared state is $\sigma_{AB}^{\prime \prime}$. In Fig.(\ref{r_P_WOW}) and  Fig.(\ref{r_AP_WOW}) we plot the respective key rates  with the strength of environmental interaction, $D$. 
From these two figures it is clear that key rate when both systems interact with environment  
is lower in comparison with the key rate when a single system interacts with environment, for
all value of the interaction strength $D$. It is worth recounting here that for the amplitude 
damping channel, the teleportation fidelity can be improved between two parties when both of
them are made to interact with the environment~\cite{Fid_enhan, Fid_Som, Fid_Tan}. Such a 
phenomenon occurs  because the decoherence effect on both systems can improve the classical 
correlation between them, enhancing in turn the teleportation fidelity.  However, no
such effect occurs for the quantum key rate which is associated with the quantum correlation 
of steerability.

\section{Improvement of  quantum key rate  using the technique of weak measurement and its reversal}

It is already known that the technique of weak measurement and its reversal can reduce the 
environmental effect modeled by ADC, i.e., it helps to protect quantum 
correlations~\cite{Fid_Tan, Weak_M_1, Weak_M_2, Weak_M_3, Weak_M_4, Weak_M_5}. In the case of 
preservation of teleportation fidelity~\cite{Fid_Tan}, both classical correlation and quantum
 correlations are involved. In the present work, we discuss the preservation of quantum 
correlations in the form of steerability with the help of the technique of weak measurement 
and its reversal. Similar to the above section, here, we consider two cases, {\it Case I} where environment affects  the system $B$ at time of traversal and {\it Case II} where environment 
affects  both systems.

{\it Case I.} To protect against the decoherence effect, Alice makes a weak measurement with 
strength $p_2$ on the system $B$ and considers the case when the system $B$ is not detected. 
In this case, depending upon the chosen initial state  $\rho^{\pm}_W$ or $\sigma^{\pm}_W$ the 
combined 
state of the systems $A$ and $B$ either becomes
\begin{eqnarray}
\rho_W = (I\otimes M_{2,0})\,\rho_{AB}^{\pm}\,(I\otimes M_{2,0}^{\dagger}),
\end{eqnarray}
 or becomes
\begin{eqnarray}
\sigma_W=(I\otimes M_{2,0})\,\sigma_{AB}^{\pm}\,(I\otimes M_{2,0}^{\dagger}),
\end{eqnarray}
where $M_{2,0}$ is defined in  Eq.~(\ref{Weak_Rev}) and $\rho_W$ ($\sigma_W$) is unnormalized. 
When the system $B$ is detected, Alice discards the state. Hence, the success probability of 
generating the state $\rho_W$ ($\sigma_W$) is given by $Tr\left[\rho_W\right]=Tr\left[\sigma_W\right]=1-D_2/2$. Next, Alice  sends the system $B$ to Bob through the environment. Due to 
environmental interaction via ADC, the shared state  becomes either
\begin{eqnarray}
\rho_E= (I\otimes W_{2,0})\,\rho_W\,(I\otimes W_{2,0}^{\dagger})\nonumber \\
+(I\otimes W_{2,1})\,\rho_W\,(I\otimes W_{2,1}^{\dagger}) 
\end{eqnarray}
or 
\begin{eqnarray}
\sigma_E= (I\otimes W_{2,0})\,\sigma_W\,(I\otimes W_{2,0}^{\dagger})\nonumber \\
+(I\otimes W_{2,1})\,\sigma_W\,(I\otimes W_{2,1}^{\dagger}).
\end{eqnarray}
After receiving the system $B$, Bob applies reverse weak measurement with the Kraus operator given by Eq.(\ref{Rev_Weak}). The final shared sate becomes either
\begin{eqnarray}
\rho_R=(I\otimes N_{2,0})\,\rho_{E}\,(I\otimes N_{2,0}^{\dagger}),
\end{eqnarray}
or
\begin{eqnarray}
\sigma_R=(I\otimes N_{2,0})\,\sigma_{E}\,(I\otimes N_{2,0}^{\dagger}).
\end{eqnarray}
Bob chooses the strength of the reverse weak measurement $q_2$ such that it maximizes the 
violation of the steering inequality~(\ref{St_FUR}) and hence, it maximizes the key rate given by Eq.(\ref{Key_Id}).

\begin{figure}[!ht]
\resizebox{8cm}{5.5cm}{\includegraphics{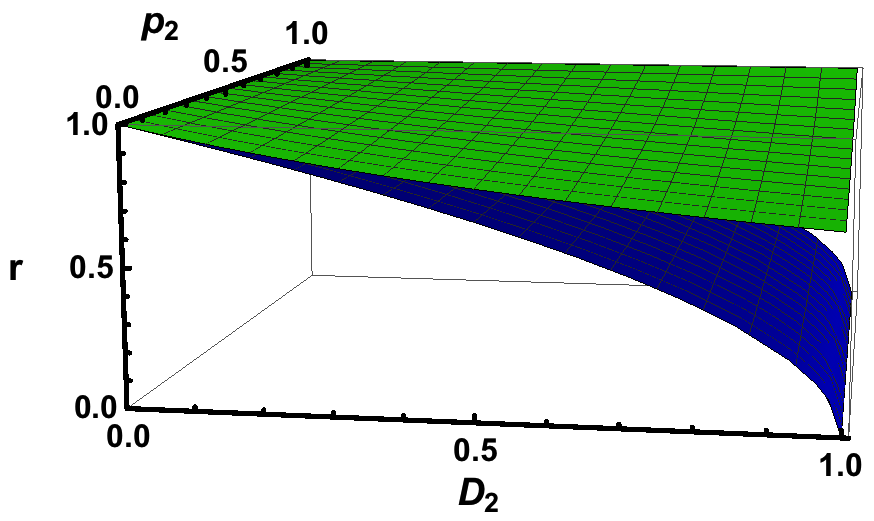}}
\caption{\footnotesize The lower bound of key rate is plotted against the strength of 
decoherence $D_2=D$ ($x$-axis) and  the strength of weak measurement  $p_2$ 
($y$-axis). The upper surface is for the key rate given by Eq.(\ref{Key_P_WR_B}) using 
the technique of weak mesurement, 
and the lower one is for the key rate given by Eq.(\ref{K_CI}).
}
\label{R_P_WR_B}
\end{figure}

When Alice prepares systems $A$ and $B$ either in the state $|\psi^{\pm}\rangle$ or in 
the state $|\phi^{\pm}\rangle$ given by Eq.(\ref{State_Ini}), the left-hand side of the inequality~(\ref{St_FUR})
 becomes 
\begin{eqnarray}
\frac{1}{2}\left[P(b_{\sigma_z}|a_{\mathcal{A}_1}) + P(b_{\sigma_x}|a_{\mathcal{A}_2})\right]=\frac{3}{4}+\frac{3}{4\sqrt{1+D_2-D_2 p_2}},
\label{St_P_WR_B}
\end{eqnarray}
where Alice's measurement settings are the same as the measurement settings when the technique 
of weak measurement is not applied, i.e., $\{\sigma_z,\,\sigma_x\}$, and the optimal strength of the reverse weak measurement is given by $q_2^O=\frac{2\,D_2\,+\,p_2\,-\,2\,D_2\,p_2}{1+D_2-D_2p_2}$. The lower bound of key rate is given by
\begin{eqnarray}
r_P^B = \log_2\left[\frac{\frac{3}{4}+\frac{3}{4\sqrt{1+D_2-D_2 p_2}}}{\frac{3}{4}-\frac{3}{4\sqrt{1+D_2-D_2 p_2}}}\right],
\label{Key_P_WR_B}
\end{eqnarray}
where the success probability of acheiving this bound is the probability of sharing the state $\rho^{R}$, i.e., $Tr[\rho_{R}]=(1-D_2)\,(1-p_2)$. In Fig.(\ref{R_P_WR_B}), we compare the 
key rate~(\ref{K_CI}) without using the technique of weak measurement with the key rate~(\ref{Key_P_WR_B}) when the technique of weak measurement and its reversal is used. From this figure 
it is clear that one can protect steerability and hence, the key rate of 1s-DIQKD from 
decoherence modeled by ADC.

{\it Case II.:} Here both the systems $A$ and $B$ interact with the environment via ADC. 
To protect the correlation from decoherence, Alice makes weak measurements on both systems. When both systems are not detected, the combined state of system $A$ and $B$ is either given by
\begin{eqnarray}
\rho_W^{\prime} = (M_{1,0}\otimes M_{2,0})\,\rho_{AB}^{\pm}\,(M_{1,0}\otimes M_{2,0}^{\dagger}),
\end{eqnarray}
 or 
\begin{eqnarray}
\sigma_W^{\prime}=(M_{1,0}\otimes M_{2,0})\,\sigma_{AB}^{\pm}\,(M_{1,0}\otimes M_{2,0}^{\dagger}).
\end{eqnarray}
Next, Alice sends the system $B$ to Bob and allows both systems to interact with the 
environment. Due to the environmental effect, the shared state  becomes either
\begin{eqnarray}
\rho_E^{\prime}= (W_{1,0}\otimes W_{2,0})\,\rho_W^{\prime}\,(W_{1,0}^{\dagger}\otimes W_{2,0}^{\dagger})\nonumber \\
+(W_{1,0}\otimes W_{2,1})\,\rho_W^{\prime}\,(W_{1,0}^{\dagger}\otimes W_{2,1}^{\dagger}) \nonumber \\
+ (W_{1,1}\otimes W_{2,0})\,\rho_W^{\prime}\,(W_{1,1}^{\dagger}\otimes W_{2,0}^{\dagger})\nonumber \\
+(W_{1,1}\otimes W_{2,1})\,\rho_W^{\prime}\,(W_{1,1}^{\dagger}\otimes W_{2,1}^{\dagger})
\end{eqnarray}
or 
\begin{eqnarray}
\sigma_E^{\prime}= (W_{1,0}\otimes W_{2,0})\,\sigma_W^{\prime}\,(W_{1,0}^{\dagger}\otimes W_{2,0}^{\dagger})\nonumber \\
+(W_{1,0}\otimes W_{2,1})\,\sigma_W^{\prime}\,(W_{1,0}^{\dagger}\otimes W_{2,1}^{\dagger}) \nonumber \\
+ (W_{1,1}\otimes W_{2,0})\,\sigma_W^{\prime}\,(W_{1,1}^{\dagger}\otimes W_{2,0}^{\dagger})\nonumber \\
+(W_{1,1}\otimes W_{2,1})\,\sigma_W^{\prime}\,(W_{1,1}^{\dagger}\otimes W_{2,1}^{\dagger}).
\end{eqnarray}
At the end, both apply reverse weak measurement having Kraus representation given by Eq.(\ref{Rev_Weak}). The final shared state either becomes
\begin{eqnarray}
\rho_R^{\prime}=(N_{1,0}\otimes N_{2,0})\,\rho_{E}^{\prime}\,(N_{1,0}\otimes N_{2,0}^{\dagger}),
\end{eqnarray}
or
\begin{eqnarray}
\sigma_R^{\prime}=(N_{1,0}\otimes N_{2,0})\,\sigma_{E}^{\prime}\,(N_{1,0}\otimes N_{2,0}^{\dagger}).
\end{eqnarray}

\begin{figure}[!ht]
\resizebox{8cm}{5.5cm}{\includegraphics{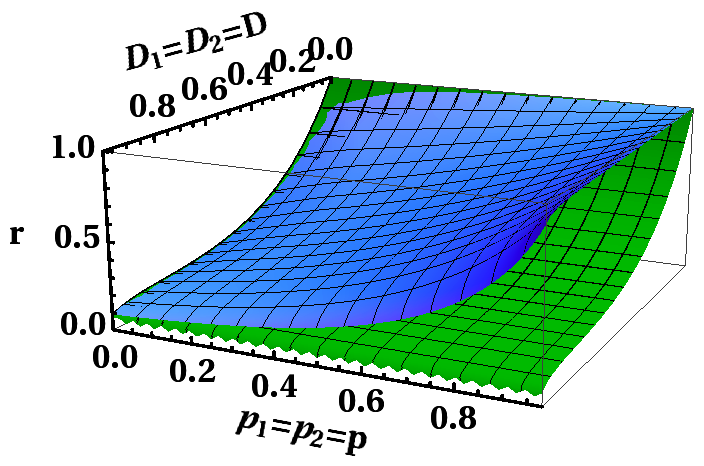}}
\caption{\footnotesize  The lower bound of key rate is plotted against the strength of 
decoherence $D_1=D_2=D$ ($x$-axis) and  the strength of weak measurement  $p_1=p_2=p$ 
($y$-axis). The upper surface is for the key rate is for the state $|\psi^{\pm}\rangle$ 
given by  Eq.(\ref{State_Ini}),  using 
the technique of weak mesurement, 
and the lower one is for the key rate given by Eq.(\ref{K_CI}).
}
\label{R_P_WR_AB}
\end{figure}

Now, we study the steerability of the shared states $\rho_{R}^{\prime}$ ($\sigma_{R}^{\prime}$). When Alice prepares systems $A$ and $B$ in the state $|\psi^{\pm}\rangle$, the choice of the set of observables for Alice is the same as the settings used when both systems interact 
with environment and the technique of weak measurements is not applied, i.e., $\{\mathcal{A}_1=\sigma_z,\,\mathcal{A}_2=\sqrt{1-D^2}\,\sigma_x-D \sigma_z\}$ . Here, for simplicity, we consider both systems interact with equal strength
with the environment, i.e., $D_1=D_2=D$, and the same strength of weak measurement and
reverse weak measurement is
used for both systems, i.e., $p_1=p_2=p$ and $q_1=q_2=q$.  We numerically maximize the 
quantity $\frac{1}{2}\left[P(b_{\sigma_z}|a_{\mathcal{A}_1}) + P(b_{\sigma_x}|a_{\mathcal{A}_2})\right]$ with respect to the strength of the reverse weak measurement $q$. In Fig.(\ref{R_P_WR_AB}), we display the improvement of key rate when Alice prepares the systems in the state 
$|\psi^{\pm}\rangle$. For the prepared state $|\phi^{\pm}\rangle$, the improvement of the key
 rate using weak measurement is not possible.

\begin{figure}[!ht]
\resizebox{8cm}{5.5cm}{\includegraphics{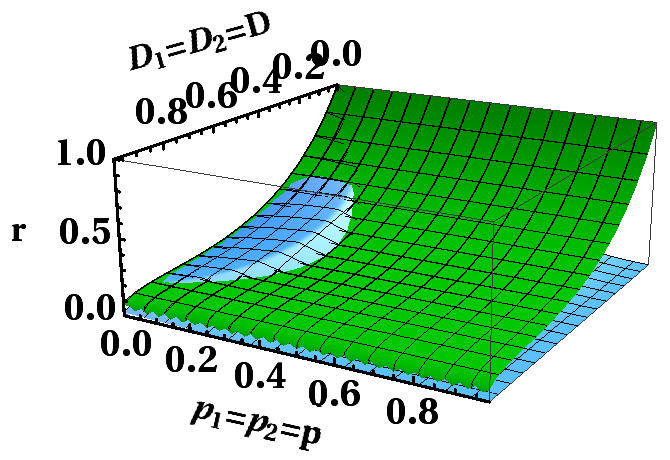}}
\caption{\footnotesize  The average key rate is plotted against the strength of 
decoherence $D_1=D_2=D$ ($x$-axis) and  the strength of weak measurement  $p_1=p_2=p$ 
($y$-axis). The surface corresponding to the case without weak measurement is the same
as in Fig.(\ref{R_P_WR_AB}). It is seen that the improvement of the average key rate
is possible for a range of values of the strength of decoherence and weak measurement.
}
\label{R_P_av_WR_AB}
\end{figure}

As this technique is associated with weak measurement, the success probability of sharing the final state $\rho_{R}^{\prime}$ ($\sigma_{R}^{\prime}$) is given by $\max_{q}\left[Tr[\rho_{R}^{\prime}]\right]$ ($\max_{q}\left[Tr[\sigma_R^{\prime}]\right]$).
Hence finally, we calculate the average steerability of the state $\rho_R^{\prime}$, where 
the average is taken over the success probability of sharing the state $\rho_R^{\prime}$.
Thus, the relevant quantity which provides the lower bound to the average key rate
depending upon the success probability is given by $\max_{q}\left[Tr[\rho_R^{\prime}]\right]\,\max_{q}\left[\frac{1}{2}\left[P(b_{\sigma_z}|a_{\mathcal{A}_1}) + P(b_{\sigma_x}|a_{\mathcal{A}_2})\right]\right]\,+\,(1-\max_{q}\left[Tr[\rho_R^{\prime}]\right]) \frac{3}{4}$, where $3/4$ is the upper bound of the steering relation (\ref{St_FUR}), achievable 
with the help of an LHS model.
Fig.~(\ref{R_P_av_WR_AB}) shows the improvement in  the average key rate corresponding to
the shared state $\rho_R^{\prime}$. We see that the technique of weak measurement allows
improvement of the average key rate in a noteable region of parameter space. 
 
\section{Conclusions}

To summarize, in the present work we have discussed the effect on steerability and the key 
rate of the 1s-DIQKD protocol when the system possessed by one or both the parties interact 
with the environment through an amplitude damping channel. It is known from earlier
studies that for a particular set of bipartite states having teleportation fidelity just 
below the quantum region, amplitude damping decoherence can improve the teleportation 
fidelity above the quantum region~\cite{Fid_enhan, Fid_Som}. This happens due to the
enhancement of classical correlations as a consequence of environmental interaction on 
both the parties. However, we show here that amplitude damping decoherence is unable to 
improve the key rate whose upper bound is fixed by steerability of bipartite states,
a quantum correlation that  falls down with the strength of interaction with the environment.

Next, we use the technique of weak measurement to protect quantum steerability and the key 
rate in the presence of amplitude damping decoherence. We show that when one of the 
parties of a
bipartite system interacts with the environment, one can protect the secret key rate in 
1s-DIQKD with the help of weak measurement and its reversal for any maximally entangled 
state. However, when both  systems interact with environment, the technique of weak 
measurement can protect the key rate only for prepared states of the 
type $|\psi^{\pm}\rangle$. Similar to the case of improvement of the teleportation 
fidelity~(\cite{Fid_Tan}), the technique of weak measurement fails to protect the key 
rate for prepared states of the type $|\phi^{\pm}\rangle$. The technique of weak 
measurement is associated 
with a success probability, as it is implemented with post-selection discarding the 
state when it is detected. We further
show here, that considering even the unsuccessful attempts (when the systems are discarded), 
the average key rate turns out to be  greater than the case where the weak measurement 
technique is not applied, for a considerable range of the interaction parameters.

{\it Acknowledgements}: ASM and SG acknowledge support from the project SR/S2/LOP-08/2013
of DST, India.

\end{document}